\def\upd{{\rm d}}

\documentclass[prl,showpacs,nofootinbib,twocolumn]{revtex4}
\usepackage{graphicx}

\begin{document}

\title{Invariant quantities in shear flow}

\author{A. Baule\footnote{Present affiliation: The Rockefeller University, 1230 York Avenue, New York, NY 10065, USA. E-mail: abaule@rockefeller.edu.}
 and R. M. L. Evans}

\affiliation{School of Physics and Astronomy, University of Leeds, Leeds LS2 9JT, United Kingdom}

\begin{abstract}

The dynamics of systems out of thermal equilibrium is usually treated on a case-by-case basis without knowledge of fundamental and universal principles. We address this problem for a class of driven steady states, namely those mechanically driven at the boundaries such as complex fluids under shear. From a nonequilibrium counterpart to detailed balance (NCDB) we derive a remarkably simple set of invariant quantities which remain unchanged when the system is driven. These new nonequilibrium relations are both exact and valid arbitrarily far from equilibrium. Furthermore, they enable the systematic calculation of transition rates in driven systems with state-spaces of arbitrary connectivity.

\end{abstract}


\maketitle

Despite the abundance of nonequilibrium states of matter in nature, a fundamental understanding of the variety of phenomena exhibited in such states is still far from complete. There have been numerous recent theoretical advances in non-equilibrium statistical mechanics, with a large body of literature devoted to establishing elegant methods for calculating the macroscopic behaviour of stochastic systems, given their microscopic transition rates, and demonstrating the efficacy of those methods by application to idealized models \cite{Privman97,Evans02}. Instead, we address the prior question of how to choose the microscopic transition rates in the first place, when defining a non-equilibrium system. The simplest generalization of thermal equilibrium, namely nonequilibrium steady states, have many features in common with equilibrium steady states, and a mathematical description starting from first principles is possible in both cases. For a common class of systems --- complex fluids in continuous shear flow --- the nonequilibrium noise imposes certain rules on the rates, akin to (but different from) the equilibrium principle of detailed balance. In this letter we derive a remarkably simple set of invariant quantities valid in steady states of sheared fluids. On the basis of these new exact nonequilibrium relations we are able to formulate a systematic method to determine the driven microscopic dynamics in state spaces of arbitrary connectivity. 

We consider a probabilistic treatment of the system under consideration (a complex fluid) in which the dynamical evolution is described in terms of a set of transition rates $\{\Omega_{ij}\}$ expressing the conditional probability per unit time to perform a transition between two microstates $i$ and $j$, where $i,j\in\{1,...,n\}$. We are interested in the properties of the system when it has settled into a nonequilibrium stationary state under the influence of continuous external driving at some distant boundaries. In the absence of driving, the set of transition rates is constrained by the principle of detailed balance, stating that, for every pair of microstates $i$ and $j$,
\begin{eqnarray}
\label{N_canDB}
	\frac{\Omega_{ij}}{\Omega_{ji}}=e^{-\beta(E_j-E_i)},
\end{eqnarray}
where $E_i$ is the microstate's energy and $\beta$ the reciprocal temperature parameter.
Detailed balance results from four fundamental properties characterizing system and heat reservoir \cite{VanKampen}: (i) ergodicity, (ii) microscopic reversibility, (iii) stationarity and (iv) conservation of energy. For a system with $\mathcal{N}$ transition rates there are $\mathcal{N}/2$ constraints in the form of Eq.~(\ref{N_canDB}) resulting from the statistics of the heat reservoir and ensuring equilibrium properties. 

Away from equilibrium, detailed balance is generally violated. In fact, the violation of Eq.~(\ref{N_canDB}) is often considered to be the defining property of a nonequilibrium state. It was recently shown \cite{EvansR04,EvansR05,Simha08} that, for a particular class of steady states including complex fluids under shear, a nonequilibrium counterpart to equilibrium detailed balance (NCDB) can be derived. The crucial observation here is that, under continuous shear, a volume element in the bulk of the fluid has the same equations of motion as at equilibrium; its Hamiltonian is not altered by the driving, but the stochastic influence of the surrounding fluid is no longer that of an equilibrium heat bath --- the reservoir is itself under shear. As a result the same properties (i)---(iv) apply as in equilibrium, amended only by an additional conserved quantity: the net average shear rate of all the fluid elements. Following either information-theoretic \cite{EvansR04,EvansR05} or Gibbsian \cite{Simha08} arguments this small modification to the ensemble results in an exact one-to-one mapping between the transition rates at equilibrium and those in the sheared steady state. That ono-to-one mapping implies a set of constraints on the  transition rates in the driven ensemble.

For continuous-time dynamics, the one-to-one mapping of NCDB is expressed as \cite{EvansR05}
\begin{eqnarray}
\label{N_DDB}
	\Omega_{ij}(\nu)=\omega_{ij}\,e^{\nu \Delta x_{ji}+\Delta q_{ji}(\nu)}.
\end{eqnarray}
Here, $\{\Omega_{ij}(\nu)\}$ denotes the set of transition rates in the sheared ensemble, parametrized by $\nu$, a Lagrange multiplier characterizing the driving strength of the reservoir. The $\omega_{ij}$ are the associated equilibrium transition rates that satisfy detailed balance, Eq.~(\ref{N_canDB}). Under the influence of shear forces applied at the boundaries of the whole ensemble, the rates are thus enhanced or attenuated with respect to equilibrium. The factor $e^{\nu \Delta x_{ji}}$ exceeds unity if the transition $i\rightarrow j$ involves a conformational change that increments the shear strain by a positive amount $\Delta x_{ij}$. This factor simply boosts every transition in the forward direction irrespective of the state space structure. By itself, it would represent a simple mean-field expression for the driven transition rates. Important non-mean-field information about the global properties of the state space is contained in the quantities $\Delta q_{ji}(\nu)$ which can be formally defined \cite{EvansR05} as
\begin{eqnarray}
\label{N_q_def}
	\Delta q_{ji}(\nu)\equiv\lim_{\tau\rightarrow\infty}\ln\left[\frac{\sum_{r=-\infty}^\infty
	G_j(r,\tau)e^{\nu r}}{\sum_{r=-\infty}^\infty G_i(r,\tau)e^{\nu r}}\right].
\end{eqnarray}
The $\{G_i(r,\tau)\} $ denote the equilibrium Green's functions, i.e., the probability that the equilibrium system, given it is in state $i$, exhibits shear strain $r$ in time $\tau$. The quantities $\Delta q_{ji}$ thus provide a measure for the increase (or decrease) in probability that the system exhibits the required shear rate (specified by $\nu$) \textit{at equilibrium} if it performs the transition $i\rightarrow j$. Therefore the mapping Eq.~(\ref{N_DDB}) expresses the fact that the likelihood of a transition not only depends on the immediate flux contribution, but also on the new microstate's prospect for future flux. It is in this sense that NCDB goes significantly beyond simple mean-field expressions for the driven transition rates. However, the crucial ingredient in the mapping, namely the set $\{\Delta q_{ji}\}$, requires knowledge of all equilibrium Green's functions of the system. This requirement has hitherto impaired the practical applicability of the theory. Below we demonstrate that the $\{\Delta q_{ji}\}$ are in fact intrinsically related to the structure of the state space and can be determined following a systematic set of network rules. We furthermore eliminate all $\Delta q_{ji}$ values to derive some remarkably simple, experimentally testable relations between the transition rates.

To this end we first introduce a graphical representation of the state space following \cite{Schnakenberg76}. Here, vertices are assigned to the different states $i$ of the system and edges to the possible transitions. If a transition is physically allowed to take place, i.e. $\omega_{ij}>0$, then equilibrium detailed balance demands that the reverse rate $\omega_{ji}$ is also non-zero. Only connected graphs are considered, since they cover most physically relevant situations. In order to guarantee that the system can exhibit a macroscopic steady state, we require the state space to have a periodic structure along a coordinate $x$ that measures shear strain.

\begin{figure}
\begin{center}
\includegraphics[width=6cm]{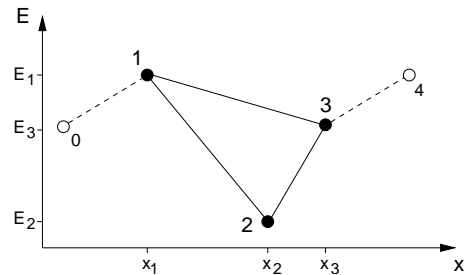}
\caption{\label{N_Fig_Graph}Example of a basic graph for a three-state system. The dotted line denotes an exterior edge connecting states in different periods. Here, states $4$ and $0$ are of the same types as $1$ and $3$ respectively. $E$ is energy and $x$ is shear strain.}
\end{center}
\end{figure}

We define the \textit{basic graph} as the graph corresponding to the non-periodic set of $n$ distinct vertices. Its set of edges will be denoted \textit{interior} edges in order to distinguish them from \textit{exterior} edges connecting vertices of the basic graph with vertices of the next or preceding period. See Fig.~(\ref{N_Fig_Graph}) for a depiction of such a basic graph. The total number of transition rates in the system is $\sum_{i=1}^{n}d_i$, where $d_i$ is the degree (or connectivity) of the $i$th vertex of the basic graph including exterior edges. The minimal number of transition rates in an $n$-state driven system is $2n$, corresponding to a graph in the form of a simple connected path. For this class of state spaces the problem of finding the driven transition rates has a particularly straightforward solution due to the presence of invariant quantities in the steady state, derived as follows.

Consider a system in microstate $i$, where it remains for a random duration $t$ until making a transition to one of the $d_i$ connected states. For a continuous-time Markov chain this waiting time has the exponential distribution $h_i(t)=\sigma_i\,e^{-\sigma_i t}$, where the total exit rate is defined as $\sigma_i\equiv\sum_{\{j\}}\omega_{ij}$. The probability that the particle jumps to site $j$ is then $P_{ij}=\omega_{ij}/\sigma_i$. The Green's function for state $i$ can be determined as follows. From state $i$ the system can only perform a transition to one of the connected states $j$; its subsequent displacement is then determined by the Green's function of state $j$. $G_i$ is therefore related to the $d_i$ Green's functions of the neighbouring sites by
\begin{eqnarray}
\label{N_greens}
	G_i(r,\tau)=C_i\int_0^\tau \upd t\, h_i(\tau-t)\sum_{\{j\}}P_{ij}G_j(r-\Delta x_{ji},t),
\end{eqnarray}
where the sum is taken over the set of states $\{j\}$ connected with $i$. For normalized Green's functions, the coefficient is $C_i=1/(1-e^{-\sigma_i\tau})$. We next transform the Green's functions to the corresponding quantities
\begin{eqnarray}
\label{N_m_def}
	m_i(\nu,\tau)\equiv\ln\sum_{r=-\infty}^\infty G_i(r,\tau)e^{\nu r},
\end{eqnarray}
whereby the difference $m_j-m_i$ is ultimately related to the $\Delta q_{ji}$ in the long time limit (cf. Eq.~(\ref{N_q_def})). Multiplying Eq.~(\ref{N_greens}) by $e^{\nu r}$ and summing over the displacement $r$ from state $i$ yields, after a shift in the summation variable,
\begin{eqnarray}
	e^{m_i(\nu,\tau)}=C_i\int_0^\tau 
	\upd t\,h_i(\tau-t)\sum_{\{j\}}P_{ij}\,e^{m_j(\nu,t)+\nu\Delta x_{ji}}.
\end{eqnarray}
Substituting for $C_i$, $P_{ij}$ and $h_i(t)$, and rearranging yields
\begin{eqnarray}
	&&(1-e^{-\sigma_i\tau})\,e^{m_i(\nu,\tau)+\sigma_i\tau}\nonumber\\
	&&=\int_0^\tau \upd t\,e^{\sigma_i t}\sum_{\{j\}}\omega_{ij}\,e^{m_j(\nu,t)+\nu\Delta x_{ji}}.
\end{eqnarray}
The integral can be removed by taking the derivative with respect to $\tau$ on both sides, eventually leading to
\begin{eqnarray}
\label{N_Gmtau}
	&&\frac{\partial}{\partial \tau}m_i(\nu,\tau)+\sigma_i-\frac{\partial}{\partial
	\tau}m_i(\nu,\tau)e^{-\sigma_i \tau}\nonumber\\
	&&=\sum_{\{j\}}\omega_{ij}\,e^{m_j(\nu,\tau)-m_i(\nu,\tau)+\nu\Delta x_{ji}}.
\end{eqnarray}
In order to ensure steady state properties we have to consider the limit $\tau\rightarrow\infty$ where $m_j-m_i\rightarrow \Delta q_{ji}$ and
\begin{eqnarray}
Q(\nu)\equiv \lim_{\tau\rightarrow \infty}\frac{\partial}{\partial \tau}m_i(\nu,\tau),
\end{eqnarray}
i.e., in the long time limit the rate of change of $m_i(\nu,\tau)$ converges to a 
state-independent function of $\nu$, the \textit{flux potential} $Q(\nu)$ \cite{EvansR05}. In Eq.~(\ref{N_Gmtau}), the result is
\begin{eqnarray}
	Q(\nu)+\sigma_i=\sum_{j}\omega_{ij}\,e^{\Delta q_{ji}(\nu)+\nu\Delta x_{ji}}.
\end{eqnarray}
On the right-hand side we can identify the transition rates in the driven steady state according to Eq.~(\ref{N_DDB}). We therefore obtain a fundamental relationship between the equilibrium transition rates, the corresponding rates in the driven steady state, and the flux potential $Q(\nu)$:
\begin{eqnarray}
\label{N_Qpi}
	Q(\nu)=\Sigma_i(\nu)-\sigma_i,
\end{eqnarray}
where $\Sigma_i(\nu)\equiv\sum_{\{j\}}\Omega_{ij}(\nu)$. Eq.~(\ref{N_Qpi}) states that, {\em for every microstate $i$, the total exit rate in the driven ensemble differs from its equilibrium counterpart by a flux-dependent but microstate-independent constant}. On the basis of this central result a number of important implications of NCDB can be derived. It turns out that it is not necessary to explicitly calculate the Green's functions of the system in order to determine the $\Delta q$'s and the driven transition rates. Rather, as we will see more explicitly below, the quantities of the NCDB formalism are intrinsically related to the graph structure via Eq.~(\ref{N_Qpi}).

It is now straightforward to formulate two sets of \textit{invariant quantities} for the steady state. The first was found previously for continuous-time dynamics \cite{EvansR05} and is a consequence of the antisymmetries $\Delta q_{ji}=-\Delta q_{ij}$ and $\Delta x_{ji}=-\Delta x_{ij}$. From Eq.~(\ref{N_DDB}) the \textit{`product constraint'} directly follows:
\begin{eqnarray}
\label{N_pcons}
	\Omega_{ij}\Omega_{ji}=\omega_{ij}\omega_{ji}.
\end{eqnarray}
Secondly, Eq.~(\ref{N_Qpi}) implies the \textit{`exit rate constraint'}:
\begin{eqnarray}
\label{N_qcons}
	\Sigma_i -\Sigma_j=\sigma_i-\sigma_j.
\end{eqnarray}
NCDB thus predicts that the product of forward and reverse transition rates and the difference of total exit rates for every pair of microstates are the same as in equilibrium and therefore invariant with respect to driving. No near-equilibrium assumptions have been made in the derivation, so the above relations are both exact and valid arbitrarily far from equilibrium.

As well as being important and elegant in their own right, Eqs.~(\ref{N_pcons}) and (\ref{N_qcons}) can be used to find the rates $\Omega_{ij}$ (Eq.~(\ref{N_DDB})) without evaluating the RHS of Eq.~(\ref{N_q_def}), as follows.
Considering the whole basic graph with $n$ states, there are $n$ equations in the form of Eq.~(\ref{N_Qpi}). This set of equations is sufficient to determine the unknown $\Delta q_{ji}$'s in Eq.~(\ref{N_DDB}). Due to the relationship $\Delta q_{ij}=-\Delta q_{ji}$, every edge of the basic graph is associated with two transition rates (forward and backward transitions) that depend on just one $\Delta q$. The number of independent $\Delta q$'s is further constrained by closed paths in the graph, i.e. paths that begin and end at the same type of vertex, since the sum of $\Delta q$'s along such a path vanishes (a \textit{`loop constraint'}) due to Eq.~(\ref{N_q_def}). The total number of independent $\Delta q$'s in the basic graph is always $n-1$, as is seem by considering first the simplest basic graph, namely all $n$ states connected as a simple path without any loops. In this case the number of edges is trivially $n$. Since one loop constraint is generated by the periodicity, there are $n-1$ independent $\Delta q$'s. From this simply connected graph all graphs of higher degree are generated by adding new edges. But adding an edge generates a new $\Delta q$ and at the same time a new loop constraint, so that the number of independent $\Delta q$'s always remains $n-1$. With this knowledge, we can formulate the following \textit{network rules} for the calculation of the driven transition rates in networks of arbitrary connectivity:

\textit{Edge rule}. Every interior edge and every pair of exterior edges in the basic graph corresponds to two rates that depend on one $\Delta q$ in Eq.~(\ref{N_DDB}).

\textit{Vertex rule}. For every vertex in the basic graph the difference between the driven and equilibrium total exit rates equals the flux potential $Q$ (Eq.~(\ref{N_Qpi}))

\textit{Loop rule}. For every closed path of edges the sum of the $\Delta q$'s along this path is zero.

In this formulation there are in total $n$ equations and $n$ unknowns, namely one $Q$ and $n-1$ $\Delta q$'s. The number of independent equations can always be further reduced by eliminating $Q$, such that one is essentially left with $n-1$ equations for $n-1$ unknown $\Delta q$'s. The solution of this system of equations fully specifies all the driven rates in the system as well as the imposed shear rate $J$, as functions of the flux conjugate parameter $\nu$ which is related to the flux via $\upd Q/\upd\nu=J$ \cite{EvansR05}.

Alternatively, an even simpler, $\nu$-independent solution can be found from Eqs.~(\ref{N_pcons}) and (\ref{N_qcons}) alone. However, the number of constraints is then not sufficient to determine all the driven rates for every network structure. On the one hand there are $\sum_{i=1}^{n}d_i/2$ product constraints and $n-1$ exit rate constraints. On the other hand, for an arbitrary graph configuration, there are $\sum_{i=1}^{n}d_i$ transition rates. Therefore only for graphs with the topology of a simple connected path (where $d_i=2$), we can completely determine the rates using only the invariant quantities without applying the network rules. In this case we have $2n$ transition rates and $2n-1$ constraints stemming from the exact relations. The transition rates are fully determined if additionally the relationship between the transition rates and the current is provided. In this formulation the driven rates depend on $J$ directly instead of being parametrized by $\nu$. The relationship between current and rates is further elucidated elsewhere \cite{Baule08}, but is straightforward to calculate for simple models.

\begin{figure}
\begin{center}
\includegraphics[width=7cm]{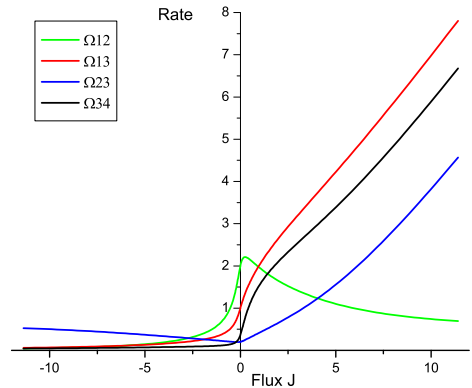} 
\caption{\label{N_Fig2}The four forward transition rates of the three-state system of Fig.~\ref{N_Fig_Graph} with $E_1=3$, $E_2=0$, and $E_3=2$ ($\beta$ set to unity). Parameter values: $\omega_{10}=\omega_{13}=1$, $\omega_{12}=2$, $\omega_{32}=1.5$, $E_2=0$, $\Delta x_{43}=0.5$, $\Delta x_{32}=1.8$, $\Delta x_{21}=1$.}
\end{center}
\end{figure}

As an illustrative example we have applied the network rules to the three state system depicted in Fig.~\ref{N_Fig_Graph}. At equilibrium the system is completely specified by the set of eight transition rates which pairwise satisfy the detailed balance condition Eq.~(\ref{N_canDB}). Let us now imagine that shear is applied to the system at the distant boundaries with a constant shear rate $J$. How does the continuous shear affect the microscopic dynamics of the system? This information is provided in a precise, unbiased way by NCDB applying Eq.~(\ref{N_DDB}). If we know the shear contribution of each individual transition at equilibrium, i.e., the set of $\{\Delta x_{ji}\}$, then the non-local quantities $\{\Delta q_{ji}\}$ can be determined via the above network rules. The results for the driven transition rates in the forward direction (positive $\Delta x$) are shown in Fig.~\ref{N_Fig2}. For forward shear three of the transition rates are strongly enhanced as expected and as would also be predicted by a mean-field model simply boosting each forward rate. However, we observe that the transition $1\rightarrow 2$ is attenuated for increased shear, indicating that the system disfavors the path via state $2$ that requires two transitions to acquire the shear increment $\Delta x_{31}$. For large shear rates the system will thus predominantly choose the direct path $1\rightarrow 3$ even though $\Omega_{12}$ also carries forward flux. A similar observation is made for the transition $\Omega_{23}$: for backward driving this transition remains significant because it connects to the favorable direct path.  

The search for fundamental principles governing the behaviour of systems in nonequilibrium situations has long been an area of intensive research. For systems maintained in a driven stationary state under the influence of a nonequilibrium heat bath, such principles can be identified in the form of NCDB, providing exact constraints on the driven transition rates arbitrarily far from equilibrium. In this letter this theory has been investigated for systems evolving in discrete state spaces. As our main result we have derived a simple relationship between the equilibrium and driven total exit rates which directly leads to a new set of invariant quantities. These invariant quantities are non-trivial predictions of NCDB for arbitrary state spaces and can be considered as new exact relations in nonequilibrium statistical mechanics.

Even in a simple three-state model, the non-local correlations of NCDB become evident. Whereas mean-field theories would simply boost a particular transition in the forward flux direction, NCDB takes into account a state's propensity for achieving flux. A forward transition can thus be attenuated if it connects to a jammed state or from where subsequent transitions carry low flux. This striking property of the theory might ultimately be able to describe the counter-intuitive phase behavior exhibited for example in real complex fluids under flow.

\acknowledgements

This work was funded by EPSRC Grant GR/T24593/01. RMLE is funded by the Royal Society.

%\bibliography{FluctuationsBib}

\end{document}